\documentclass[12pt]{article}
\usepackage{epsfig,amsfonts,amssymb}
\usepackage{hyperref}
\usepackage{cite}
\input epsf.sty
\usepackage{cite}

\def\ZZZ{{\hbox{ Z\kern-1.6mm Z}}}
\def\RRR{{\hbox{ R\kern-2.4mm R}}}
\def\CCC{{\hbox{ C\kern-2.0mm C}}}
\def\zzz{{\hbox{z\kern-1mm z}}}

\newcommand{\qeq}{{\hbox{=\kern-2.3mm ? \kern.5mm }}}
\renewcommand{\qeq}{=}

\newcommand{\eps}{\epsilon}

\newcommand{\FF}{{\cal F}}

\newcommand{\wt}{\widetilde}
\newcommand{\wh}{\widehat}

\newcommand{\be}{\begin{equation}}
\newcommand{\ee}{\end{equation}}
\newcommand{\ben}{\begin{eqnarray}\displaystyle}
\newcommand{\een}{\end{eqnarray}}

\newcommand{\refb}[1]{(\ref{#1})}

\newcommand{\sectiono}[1]{\section{#1}\setcounter{equation}{0}}

\def\one{{\hbox{ 1\kern-.8mm l}}}
\def\zero{{\hbox{ 0\kern-1.5mm 0}}}

\newcommand{\bea}[1]{\begin{eqnarray}\label{#1} }
\newcommand{\eea}{\end{eqnarray}}

\newcommand{\eqref}{\refb}

\usepackage{bm}
\usepackage[table]{xcolor}

\begin{document}

\begin{flushright}
\end{flushright}

\vskip 12pt

\baselineskip 24pt

\begin{center}
{\Large \bf  Equivalence of Two Contour Prescriptions 
in Superstring Perturbation Theory}

\end{center}

\vskip .6cm
\medskip

\vspace*{4.0ex}

\baselineskip=18pt

\centerline{\large \rm Ashoke Sen}

\vspace*{4.0ex}

\centerline{\large \it Harish-Chandra Research Institute}
\centerline{\large \it  Chhatnag Road, Jhusi,
Allahabad 211019, India}

\centerline{and}

\centerline{\large \it Homi Bhabha National Institute}
\centerline{\large \it Training School Complex, Anushakti Nagar,
    Mumbai 400085, India}

\centerline{and}

\centerline{\large \it School of Physics, 
Korea Institute for Advanced Study}
\centerline{\large \it  Seoul 130-722, Korea}

\vspace*{1.0ex}
\centerline{\small E-mail:  sen@mri.ernet.in}

\vspace*{5.0ex}

\centerline{\bf Abstract} \bigskip

Conventional superstring perturbation theory based on the world-sheet approach
gives divergent results for the S-matrix whenever the total center of mass energy of 
the incoming particles exceeds the threshold of production of any final state consistent 
with conservation laws.  Two systematic approaches have been suggested for 
dealing with this difficulty.  The first one involves deforming the integration cycles
over the moduli
space
of punctured Riemann surfaces into complexified moduli space.  The second one treats
the amplitude as a sum of superstring field theory Feynman diagrams and deforms 
the integration contours
over loop energies of the Feynman diagram into the complex plane. In this paper we
establish the equivalence of the two prescriptions to all orders in perturbation theory.
Since the second approach is known to lead to unitary amplitudes, this establishes the
consistency of the first prescription with unitarity.

\vfill \eject

\baselineskip=18pt

\tableofcontents

\sectiono{Introduction} \label{sintro}

The Polyakov formalism gives an elegant description of closed
string scattering amplitudes
by expressing the $g$-loop, $N$-point scattering amplitude as a 
single term, given by an
integral over the moduli space of genus $g$ Riemann surfaces with $N$ punctures,
with integrand given by a correlation function of the vertex operators of external states,
ghost fields and picture changing operators (for heterotic and type II strings) on the
Riemann surface. However while actually computing the integral over the moduli one
runs into difficulties: in most physical situations the result diverges. The divergence
occurs whenever the total center of mass energy of the external states is above the
threshold of production of any final state\cite{marcus,sundborg,amano}. 
Since the initial state can always appear as
the final state, this is true essentially in all cases with a few exceptions. One such
exception is  one loop
two point function of stable string states needed for computing one loop renormalized mass.

There is actually a physical reason behind this. The naive expression based on 
Polyakov formalism is always real. But in any sensible quantum field theory we expect
the amplitudes to possess an imaginary part so as to be consistent with unitarity. Therefore
in order that such an imaginary part appears, the original amplitude must diverge, and
must be defined by some sort of analytic continuation that could generate an imaginary
part. Such analytic continuation has been carried out successfully 
for one loop four point function of massless
fields\cite{9302003,9404128,9410152}. This requires breaking up the original integral
into several parts and analytic continuation of each part independently.

A general procedure for dealing with this problem was suggested in \cite{berera,1307.5124}. 
In this approach one
complexifies the moduli space of Riemann surfaces with punctures, and deforms the
integration over the modular parameters into the complexified moduli space near the
boundary of the moduli space following a suitable prescription. This in principle gives 
a procedure for extracting finite answers for the Polyakov amplitudes. Furthermore, due
to the deformation of the integration contour into the complexified moduli space,
the amplitude develops an imaginary part.

There is a second proposal for dealing with this problem based on superstring field 
theory\cite{1604.01783}.
In this approach one represents Polyakov amplitude as a sum of Feynman diagrams
of superstring field theory, with each Feynman diagram given by usual integrals over
loop momenta. The integrals over the spatial components of loop momenta are
taken to be along the real axes as usual, but integration over loop energies are chosen
to be along specific contours in the complex plane, beginning at $-i\infty$ and
ending at $i\infty$. This gives results free from divergences, but has imaginary
part due to the loop energy integrals. The added advantage of this prescription is that
this has been shown to give unitary S-matrix\cite{1604.01783,1606.03455,1607.08244}.

Our goal in the paper will be to show that the results obtained by these two apparently
different procedures are the same. 
For one loop two point function this was shown in \cite{1607.06500}, but our result holds for
all amplitudes to arbitrary order in perturbation theory.
The unitarity of the S-matrix computed in the second
approach then automatically establishes unitarity of the S-matrix computed in the first
approach.

In \S\ref{stwo} we give a detailed description of these two prescriptions for getting finite
results for string theory amplitudes. \S\ref{sequiv} proves the equivalence of these two
procedures.

\sectiono{The two contour integration prescriptions} \label{stwo}

Even though ultimately we shall be interested in the computation of on-shell amplitudes,
it will be useful to begin by discussing properties of general off-shell amplitudes using
the language of superstring field theory. First
consider an off-shell $n$-point interaction vertex of superstring field theory with external legs of
masses $m_1,\cdots m_n$ and momenta $k_1,\cdots k_n$. The general form of the
vertex is of the form
\be \label{ever1}
\int [dy] \, \exp\left[- \sum_{i,j} g_{ij}(y) k_i\cdot k_j\right]
P(y, \{k_i\})\, ,
\ee
where $y$ denote collectively 
the moduli of a Riemann surface with $n$ punctures, the integration
runs over a region of the moduli space that does not include any degenerate Riemann
surface, $g_{ij}(y)$ is some function of the moduli and $P(y, \{k_i\})$ is a function of the
moduli and polynomial in the $k_i$'s and 
depends also on the quantum numbers of the external states of the vertex. 
The integrand in \refb{ever1} is the correlation
function of off-shell vertex operators inserted at the punctures using specific local
coordinates appropriate to a particular version of the superstring field theory.
$g_{ij}(y)$ can be made positive definite by `adding
stubs'  to the vertices\cite{sonoda,9301097} -- the effect of which is to 
rescale the local coordinates at the punctures\footnote{This will have
to be accompanied by a change in the region of integration of $y$ for higher order 
vertices.} by some number $\{\lambda_i(y)\}$ and
consequently scale the correlation function by factors of 
$\prod_i (\lambda_i(y))^{k_i^2+m_i^2}$. By taking $\lambda_i(y)$ to be sufficiently small
we can add arbitrarily large diagonal components to $g_{ij}(y)$.
We can now compute contributions from Feynman diagrams using these vertices.
The propagator has the standard form $(k_i^2+m_i^2)^{-1}$ possibly multiplied by
some polynomial
in $k_i$. If we denote by $\{\ell_s\}$ the independent loop momenta, by $\{p_\alpha\}$
the external momenta and by $\{k_i\}$ the momenta carried by individual internal
propagators, given by linear combinations of $\{\ell_s\}$ and $\{p_\alpha\}$, then
the contribution to a Feynman diagram takes the general form
\ben  \label{e1}
&& \int [dY] \int \prod_{s}d^D \ell_s \, \exp\left[-G_{rs}(Y) \, \ell_r \cdot 
\ell_s - 2 H_{s\alpha}(Y) \, \ell_s \cdot p_\alpha
- K_{\alpha\beta}(Y) \, p_\alpha \cdot p_\beta\right]\nonumber \\
&& \hskip 1in \times  \prod_i (k_i^2+m_i^2)^{-1} \, Q(Y, \ell, p) \, ,
\een
where $Y$ denotes collectively all the moduli from all the vertices, and $G_{rs}$,
$H_{s\alpha}$ and $K_{\alpha\beta}$ are matrices that arise by combining the
exponential factors \refb{ever1} from all the vertices after expressing the momenta
$k_i$ carried by various propagators in terms of independent
loop momenta and external momenta.
$Q(Y, \ell, p)$ is a function of the moduli $Y$ and a polynomial in the $\ell_i$'s and $p_\alpha$'s,
arising from the products of the factors of $P$ from each vertex and the numerator
factors in various propagators.
Positive definiteness of $g_{ij}(y)$ in \refb{ever1} ensures that the matrix
$\pmatrix{G & H\cr H^T & K}$ is positive definite and hence $G$ and $K$ themselves are
positive definite.
As we shall discuss shortly, \refb{e1} is ill defined at this stage since the
integral over loop energies diverge if we take them to be along the real axes.

Before going on we note that
adding stubs to vertices
serves another purpose. By multiplying each vertex by a factor of
$(\lambda_i(y))^{m_i^2}$ for small positive constants $\lambda_i(y)$,
it suppresses contribution to the vertex from massive string
states and makes the sum over  intermediate states, whose number at some
mass level $m$ grows as
$e^{c\, m}$ for some positive constant $c$, converge. For this reason we shall
not worry about the convergence of the sum over intermediate states and
focus on the possible divergences in the contribution from fixed intermediate states.

The following formal manipulation converts \refb{e1} into the usual expression
for amplitudes as integrals over the moduli space of Riemann surfaces.
We first replace each propagator
by 
\be \label{e2}
(k_i^2+m_i^2)^{-1} = \int_0^\infty dt_i \exp[-t_i(k_i^2+m_i^2)]\, ,
\ee
ignoring the fact that $k_i^2+m_i^2$ may have negative real part and hence
the above relation may not be valid. 
For each propagator 
we also introduce an angular variable $\theta_i$ 
whose integral imposes the $L_0=\bar L_0$ constraint that the propagating states
must satisfy.\footnote{Superstring field theory actually only has states with $L_0=\bar L_0$
as propagating states, but in order to make contact with moduli space integral representation
we need to temporarily relax the constraint and allow other states in the conformal field
theory, not satisfying this constraint, to propagate. We compensate for it by introducing
the projection operator $(2\pi)^{-1} \int_0^{2\pi} d\theta_i e^{i\theta_i(L_0-\bar L_0)}$.}
In the second step we 
carry out integration over
all loop momenta by gaussian integration rules, pretending that 
the integrals converge even though the integral over loop energies may not
converge. This leads to an expression of the form
\be \label{e3pre}
\int dY \prod_j \int_0^\infty dt_j \, \int_0^{2\pi} d\theta_j\, \FF(Y, \{t_i\}, \{\theta_i\},
\{p_\alpha\})\, ,
\ee
for some function $\FF$. 
Together the integral over $Y$, $\{\theta_i\}$ and $\{t_i\}$
can be identified as integration over the moduli space of Riemann surfaces.
A given Feynman diagram
of course covers only part of the full moduli space, but when we add the contribution
from all Feynman diagrams we recover the integral over the full moduli space. For our
analysis it will be sufficient to work with an expression where integration over the angular
variables $\theta_i$ have already been performed, yielding
\be \label{e3}
\int dY \prod_j \int_0^\infty dt_j \, F(Y, \{t_i\}, \{p_\alpha\})\, ,
\ee
where $F$ is obtained from $\FF$ after performing integration over the $\theta_j$'s.

Let us now examine the original expression \refb{e1} in little more detail.
If we regard the integration over all loop momenta to be running along the real axis then
this integral diverges due to the following reason. The $\ell_i^0$ dependent
quadratic term in the exponent is given by $\exp[G_{rs}(Y) \, \ell_s^0 \, \ell_r^0]$. Since
$G_{rs}$ is positive definite, this diverges exponentially for large $\ell_s^0$, making the
integral divergent. There are also additional divergences from the subspaces on which one
or more of the $k_i^2+m_i^2$ factor vanishes.  
A different problem exists in \refb{e3} in that 
the integrals over $t_i$'s typically diverge for large $t_i$. This reflects
that the replacement \refb{e2} may not be a valid one due to the real part of
$k_i^2+m_i^2$ being negative.

There are
two approaches to this problem which we now review.
In both approaches
we shall multiply all the  external
energies by a common complex number $u$ and define the integral as the limit $u\to 1$
{\it from the first quadrant.}\footnote{In our notation first quadrant will
not include the real axis where the amplitude is expected to encounter poles and
branch cuts and have to be defined as the $u\to 1$ limit of the amplitude in the
first quadrant.} 
Therefore if the physical external momenta of the $\alpha$-th
particle is $(E_\alpha, \vec p_\alpha)$,
we set $p_\alpha=(u E_\alpha, \vec p_\alpha)$. Since all the $E_\alpha$'s are non-zero in a
physical situation, the $p_\alpha^0$'s are also non-zero as long as $u\ne 0$.
\begin{enumerate}
\item{\bf Moduli space contour integral:} In this approach 
we replace the integral \refb{e3} by\cite{berera,1307.5124}
\be \label{e4}
 \int dY \prod_j \int_0^{i\infty} dt_j \, \, F(Y, \{t_i\}, \{p_\alpha\})\, .
\ee 
We shall argue in \S\ref{sequiv} that this gives a finite result for all $u$ in the first quadrant.
This will be called the moduli space contour integral representation since from the
point of view of the world-sheet theory, $t_i$ are part of the coordinates of
the moduli space of Riemann surfaces. 
For given $\{(E_\alpha,\vec p_\alpha)\}$
we shall denote the result of this computation for general $u$ by the function 
$F_{\rm moduli}(u)$.

In the original prescription of \cite{1307.5124} the integration contour over
$t_j$ was taken to be along the real axis up to some constant value $t_0$ and
then turned towards $t_0+i\infty$. We can absorb the integration up to $t_0$
into the definition of the vertices by extending the region of integration over $y$
in \refb{ever1} by adding stubs to lower order vertices. Therefore we do not suffer
from any loss of generality in taking the $t_j$ integral from 0 to $i\infty$. The
prescription of \cite{1307.5124} also had a damping factor $e^{i\eps t_j}$ in the
integrand for each $t_j$. As we shall see,
we can dispense with this damping factor as long as we
define the amplitude as a result of taking $u\to 1$ limit from the first quadrant.
\item
{\bf Loop energy contour integral:} In this approach we work directly with the momentum
space integrals \refb{e1} without using the Schwinger parameter representation \refb{e2}.
We let the contours of integration over loop
energies begin at $-i\infty$ and end at $i\infty$ in order to make the integral converge. 
In the 
interior of the complex
plane the
 integration contours are chosen as follows\cite{1604.01783}.
If we take $u$ to be on the imaginary
axis and take the loop energy integration contours to be along the imaginary axis then
the energies carried by all the internal lines are imaginary and therefore the 
$(k_i^2+m_i^2)^{-1}$ factors in \refb{e1} do not have any poles on the integration contours.
Therefore for this configuration the position of each pole relative to the loop energy
integration contours is well defined. As we deform $u$ towards 1, the integration contours
must be deformed to maintain this relative position, i.e.\ no pole should cross the 
integration contour during this deformation. It was shown in \cite{1604.01783} 
that this is always 
possible, i.e. we never encounter a situation where during the deformation two poles
approach each other from opposite sides of the contour preventing us from deforming the
contour away from the poles. 

In the following we shall use a slight variation of this prescription where 
we take the
ends of the loop energy integration contours to approach $\pm e^{i\phi_0} \infty$, 
for some fixed angle $\phi_0$ in the range 
\be \label{ethetarange}
\pi/4 < \phi_0<\pi/2\, .
\ee
$(\ell_i^0)^2$ will still have negative real part for $\ell_i^0\to \pm e^{i
\phi_0}\infty$, making the integrals convergent. 
To fix the positions of the poles relative to the contour, 
we can start with $u= r e^{i\phi_0}$ for $0< r<\infty$. In
this case, if we take the loop energy integrals to lie along the axis 
$\ell_i^0=|\ell_i^0|e^{i\phi_0}$, then
the energies carried by all the propagators have phase proportional to $e^{i\phi_0}$ and
$k_i^2+m_i^2$ is non-zero on the contour. As we now
deform $u$ to 1, we deform the contours accordingly with the ends fixed at
$\pm e^{i\phi_0}\infty$ so that the poles lie on
same side of the integration contour as for $u=r e^{i\phi_0}$.

Since the integral converges for all $\phi_0$ in the range \refb{ethetarange}, 
we can deform
$\phi_0$ to $\pi/2$ without affecting the result of the integral. Therefore 
the new contours
give the same result as the old contours corresponding to the choice $\phi_0=\pi/2$.

The above prescription can be reexpressed by saying that we define the amplitude at
$u=1$ by analytic continuation from the line $u=r e^{i\phi_0}$ for $0<r<\infty$. Since 
during this deformation we never encounter a situation where the momentum 
integration contours are pinched by poles approaching each other
from opposite sides of the contour\cite{1604.01783}, the integral has no
singularity in the first quadrant and gives an explicit representation of the
analytically continued function. For fixed $\{(E_\alpha,\vec p_\alpha)\}$
we shall denote the result of this computation for general $u$ by the function 
$F_{\rm energy}(u)$.
\end{enumerate}

Each of these prescriptions has its own advantages. The moduli space contour
integral prescription has the advantage of
preserving the original form of the Polyakov amplitude except for the deformation
of the integration contour over the moduli into the complexified moduli space. The 
approach based on loop energy contour integral has the advantage of being formulated
in the language of superstring field theory which is needed for addressing issues
of mass renormalization and shift of vacuum under quantum correction when they occur.
Furthermore this approach leads to a proof of unitarity of perturbative superstring theory.
Clearly it will be beneficial to determine if these two approaches are equivalent. This is
the question to which we now turn. 

\sectiono{Equivalence of the two prescriptions} \label{sequiv}

Our goal will be to show that the two prescriptions give the same result. We shall show
this by establishing the following two properties of $F_{\rm moduli}(u)$:
\begin{enumerate}
\item The integration over the $t_j$'s in \refb{e4}
converges for all $u$ in the first quadrant. This shows that
$F_{\rm moduli}(u)$ given by
\refb{e4} is an analytic
function of $u$ in the first quadrant.
\item When $u=re^{i\phi_0}$, $\phi_0$ being the same angle in the range 
\refb{ethetarange} that was used in defining the
loop energy contour integral prescription, $F_{\rm moduli}(u)$ and 
$F_{\rm energy}(u)$ coincide.
\end{enumerate}
$F_{\rm energy}(u)$ was already shown to be an
analytic function of $u$ in the first quadrant of the complex $u$ plane\cite{1604.01783}. 
Since both $F_{\rm moduli}(u)$ and $F_{\rm energy}(u)$  are analytic functions
of $u$ in the first quadrant and coincide on the $u=r e^{i\phi_0}$ line,
they must agree everywhere in the first quadrant
and in particular in the $u\to 1$ limit. This is the desired result.

Therefore it remains to prove the two assertions made above. 
For this analysis, while defining $F_{\rm moduli}(u)$ 
we shall skip the steps involved in eqs.\refb{e2}-\refb{e4} by directly
making the replacement
\be \label{e5}
(k_i^2+m_i^2)^{-1}=\int_0^{i\infty} dt_i e^{-t_i (k_i^2+m_i^2)}
=\int_0^{i\infty} dt_i e^{t_i \, (k_i^0)^2 -t_i \, (\vec k_i^2+m_i^2)}
\, ,
\ee
and then carrying out integration over loop momenta by treating them as gaussian
integration. The main reason for going through step \refb{e2} was to display the relation
to the conventional expression \refb{e3pre} for the amplitudes as integrals over the moduli space
of Riemann surfaces, but this will not play any role in the analysis below since we shall
be working with the deformed integration contours.
Substituting \refb{e5} into \refb{e1} we get
\ben  \label{e11}
&& \int [dY] \prod_j \int_0^{i\infty} dt_j \int \prod_{s}d^D \ell_s \,  \Bigg[
Q(Y, \ell, p) \nonumber \\
&& \times \exp\Bigg\{-G_{rs}(Y) \ell_r \cdot 
\ell_s - 2 H_{s\alpha}(Y) \ell_s \cdot p_\alpha
- K_{\alpha\beta}(Y) p_\alpha \cdot p_\beta -\sum_i t_i (k_i^2+m_i^2) \Bigg\}\Bigg]\, .
\een
The moduli space contour integral prescription tells us that we should now carry out the
integration over the $\ell_s$'s formally regarding them as Gaussian integrals 
(after rotating $\ell_s^0$ to $i \ell_s^E$). This
gives an expression for $F_{\rm moduli}(u)$ of the form
\be \label{e12}
F_{\rm moduli}(u)=  \int [dY] \prod_j \int_0^{i\infty} dt_j  \, R(Y, p,t) \, \exp\left[
 - J_{\alpha\beta}(Y,t) p_\alpha \cdot p_\beta - \sum_i t_i m_i^2\right]\, .
 \ee
The exponent $J_{\alpha\beta}$ is obtained 
by `completing the squares' in the exponent of \refb{e11} after
expressing $k_i$'s in terms of $\{\ell_s\}$ and $\{p_\alpha\}$ 
and then picking up the left-over $\ell_s$ independent terms.
$R(Y,p,t)$ is determined by the $Q(Y,\ell,p)$ factor and 
is a polynomial in the $p_\alpha$'s and rational function 
of the $t_i$'s. 
 
The second assertion made at the beginning of this section is
easier to prove; so we shall begin with this one. We start with $F_{\rm energy}(u)$. 
As mentioned below \refb{ethetarange},
for $u=re^{i\phi_0}$,
in the computation of $F_{\rm energy}(u)$ 
we can take all the loop energies to have phase $e^{i\phi_0}$.
In this case $(k_i^0)^2$ has phase $e^{2i\phi_0}$ for each propagator making each
propagator non-singular.
The strategy will be to show that the formal steps used in going from 
$F_{\rm energy}(u)$  to $F_{\rm moduli}(u)$ now hold as true identities.
We shall begin with
\refb{e5}. First note that
\be 
t_i \,(k_i^0)^2 = i e^{2i\phi_0} |t_i|\, |(k_i^0)^2|
\ee
has negative real part proportional to $-\sin(2\phi_0)$
for $\phi_0$ in the range \refb{ethetarange}.  On the other
hand $-t_i (\vec k_i^2 +m_i^2)$ is purely imaginary and therefore just contributes a
phase to the integrand of \refb{e5}. Therefore the integration over 
$t_i$ in \refb{e5} converges and the use of \refb{e5} is justified.
The only case where the integral may fail to converge is if
$k_i^0=0$. But these are subspaces of measure zero in the
loop energy integration space, and except in low dimensions where infrared
divergences are severe, the contribution from such subspaces can be ignored.

After the replacement \refb{e5} the loop energy integration of \refb{e11}, 
taken along the
constant phase lines with phase $e^{i\phi_0}$ with $\phi_0$ in the
range \refb{ethetarange}, also converges since the
$G_{rs}(Y) \ell_r^0 \ell_s^0$ term in the exponent 
has  negative real part proportional to $\cos(2\phi_0)$ due
to positive definiteness of $G_{rs}(Y)$ and the $t_i (k_i^0)^2$ factors in the exponent
have negative real
part proportional to $-\sin(2\phi_0)$. 
Integration over spatial components of loop momenta converges due to the
factor of $e^{-G_{rs}(Y)\vec\ell_r\cdot\vec\ell_s}$, the
$e^{-t_i \vec k_i^2}$ factors giving pure phase. 
Therefore we can carry out the integration over
loop momenta treating them as gaussian integrals, leading to the expression
for $F_{\rm moduli}(u)$ given in \refb{e12}. This shows that for $u=r e^{i\phi_0}$,  
$F_{\rm moduli}(u)$ and $F_{\rm energy}(u)$ agree for all $r$ in the range
$0< r<\infty$.

We now turn to the first assertion, i.e. showing that $F_{\rm moduli}(u)$ given
in \refb{e12} is an
analytic function of $u$ in the first quadrant. For this we need to show that the 
 integrations over $\{t_i\}$ in the expression \refb{e12}
 converge as long as $u$  lies in the
 first quadrant. This in turn
requires determining the behavior of $J_{\alpha\beta}$
in the limit when some of the $|t_i|$'s become large.
 When all $|t_i|$ are large, say $t_i =i \lambda \, v_i$ with large positive 
 $\lambda$ and real 
 positive $v_i$,
 then the effect of the $G_{rs}$, $H_{s\alpha}$ and $K_{\alpha\beta}$ terms in the
 exponent of \refb{e11}
 can be neglected. In this limit $\lambda$ can be absorbed into a rescaling 
 of $\{p_\alpha\}$ and the integration variables $\{\ell_s\}$, and we have
 \be \label{eJ}
 J_{\alpha\beta}(Y, t) \simeq i \lambda \, L_{\alpha\beta}(\{v_i\}) \, .
\ee
We shall now argue that the matrix $L_{\alpha\beta}$ is positive semi-definite.  For this
we note that if we consider the integration over euclidean momenta then the same analysis
that leads us from \refb{e11} to \refb{e12}, \refb{eJ} tells us that for a real
positive parameter $\sigma$
\be
\int \prod_{s=1}^N d^{D-1}\ell_s \exp\left[- \sigma \, \sum_i v_i \vec k_i^2\right] 
\propto \sigma^{-N(D-1)/2}\exp[-\sigma \, L_{\alpha\beta}(\{v_i\}) \vec p_\alpha \cdot \vec p_\beta]\, .
\ee
This result is no longer formal, it holds as a true identity.
Now the integrand on the left hand side is real and is a monotonically  decreasing function of
$\sigma$  for all $\{\vec p_\alpha\}$. 
Therefore the right hand side must also be a monotonically decreasing
function of $\sigma$ for all $\{\vec p_\alpha\}$. This shows that 
$L_{\alpha\beta}\, \vec p_\alpha \cdot \vec 
p_\beta\ge 0$ for any $\{\vec p_\alpha\}$, i.e.\  $L_{\alpha\beta}$
is a positive semi-definite matrix.
Substituting \refb{eJ} into \refb{e12} we can now see that for generic $\{E_\alpha\}$ the
$J_{\alpha\beta}(Y,t) \, p^0_\alpha p^0_\beta
\simeq i\, \lambda \, u^2\, L_{\alpha\beta} \, E_\alpha E_\beta$ term has negative
real part proportional to $-\sin(2 \, \hbox{phase}(u))$
in the range $0<\hbox{phase}(u)<\pi/2$. Rest of the terms in the exponent 
of \refb{e12} are either  purely imaginary or subleading in the large $\lambda$ limit. 
Therefore the integrand is exponentially
suppressed in this limit and the integral converges.

Next we have to consider the case where a subset of the $t_i$'s become large keeping
the others finite.  First suppose that the $k_i$'s corresponding to these $t_i$'s can be
taken as independent loop momenta, i.e.\ there is no relation between these $k_i$'s.
In that case in the $t_i\to\infty$ limit, the (formal) 
integration over these $k_i$'s in \refb{e11}
using the rules of gaussian integration effectively
sets
them to zero. The remaining integrand is then independent of these large $t_i$'s except for
the oscillatory factors proportional to $e^{-t_i m_i^2}$, 
$t_i^{-D/2}$ multiplicative factors coming from the integration over the $k_i$'s and
possibly further negative powers of $t_i$ due to the polynomials in the
$k_i$'s coming from the $Q(Y,\ell,p)$ factor in the integrand
of \refb{e11}. After the (formal) integration over the rest of the loop momenta
are performed, we get a finite $J_{\alpha\beta}$ independent of the
large $t_i$'s. Therefore the $-J_{\alpha\beta} \,
p_\alpha\cdot p_\beta$ term in the
exponent of \refb{e12} does not provide any suppression factor. However the $t_i^{-D/2}$ factors
suppress the integrand in the 
large $t_i$ limit and give a finite integral
for $D>2$. For $D\le 2$ we anyway expect infrared 
divergences in the 
presence of massless states 
since a single massless
propagator can give divergent contributions of the form $\int d^D k/k^2$.
Therefore we shall not worry about this case.

Let us now consider the case where a subset of the $t_i$'s become large and the corresponding
$k_i$'s satisfy one of more constraints relating linear combinations of these $k_i$'s
to linear combinations of some external momenta. Let us denote by $\wh t_a$ the $t_i$'s
which become large and by $\wt t_m$ the $t_i$'s which remain finite.
We now parametrize the
$\wh t_a$'s as $\wh t_a =i\, \lambda \, \wh v_a$ where $\lambda$ is large and positive and 
$\wh v_a$'s are finite and positive. Furthermore let us denote by $\{P_s\}$ the linear
combinations of the $\{p_\alpha\}$'s which enter the constraints involving the $k_i$'s whose
Schwinger parameters are large.
In this case we can first carry out the formal 
integration over the $k_i$'s associated with
the large $t_i$'s subject to these constraints and generate a term in the exponent 
of the form $-i\,\lambda \, \wh L_{st} (\{\wh v_a\}) P_s \cdot P_t$ in addition to
a finite term, and then carry out formal integration over the rest
of the loop momenta, generating a finite term in the
exponent.
Following the same logic as the one given before one can argue that 
$\wh L$ is positive semi-definite.
The previous arguments can now be repeated to show that for generic $\{E_\alpha\}$,
$i\,\lambda\,\wh L_{st} \, P^0_s P^0_t$
has negative real part proportional to $-\sin(2\, \hbox{Phase}(u))$, 
making the integration over $\wh t_a$'s converge.

This establishes that $F_{\rm moduli}(u)$ is an analytic function of $u$ in the first quadrant.
As argued before, this in turn proves that $F_{\rm moduli}(u)=F_{\rm energy}(u)$.
 
\bigskip

{\bf Acknowledgement:}
I wish to thank Roji Pius for useful discussions.
This work  was
supported in part by the 
DAE project 12-R\&D-HRI-5.02-0303, J. C. Bose fellowship of 
the Department of Science and Technology, India and the KIAS 
distinguished professorship.

\end{document}